\documentclass[twocolumn,aps,prl,superscriptaddress]{revtex4-2}
\usepackage{amsmath}
\usepackage{graphicx}
\usepackage{MnSymbol}
\usepackage{hyperref}
\usepackage{fancyhdr}

\begin{document}

%\title{Comment on ``The ``pure-shear'' fracture test for viscoelastic elastomers and its revelation on Griffith fracture''}
\title{Some comments on the fracture of viscoelastic solids}

\author{B.N.J. Persson}
\affiliation{Peter Gr\"unberg Institute (PGI-1), Forschungszentrum J\"ulich, 52425, J\"ulich, Germany}
\affiliation{MultiscaleConsulting, Wolfshovener str. 2, 52428 J\"ulich, Germany}

\author{G. Carbone}
\affiliation{Department of Mechanics, Mathematics and Management, Politecnico di Bari, Via Orabona 4, 70100, Bari, Italy}

\author{C. Creton}
\affiliation{Soft Matter Sciences and Engineering (SIMM), ESPCI Paris, PSL University, 
Sorbonne University, CNRS, 75005 Paris, France}

\author{G. Heinrich}
\affiliation{Technische Universit\"at Dresden, Institut f\"ur Textilmaschinen und Textile
Hochleistungswerkstofftechnik, Hohe Stra{\ss}e 6, D-01069 Dresden, Germany}

\author{T. Tada}
\affiliation{Sumitomo Rubber Industries, Ltd., Material Research \& Development HQ. 2-1-1, Kobe 651-0071, Japan}

\begin{abstract}
Crack propagation in viscoelastic solids like rubber is of great practical importance.
Shrimali and Lopez-Pamies have proposed a new interesting approach for the crack propagation in viscoelastic solids. 
We give comments on the validity of the theory and point out some effects not included in the theory.
\end{abstract}

\maketitle

\thispagestyle{fancy}

%%%%%%%%%%%%%% main text %%%%%%%%%%%%%%%%

Subm. to Extreme Mechanics Letters, 1.09.2023
\vskip 0.2cm

{\bf Introduction}

Crack propagation in rubber-like materials has many applications, e.g. to tire wear\cite{Gert0,Gert00},
and has been studied for a long time\cite{Riv,Gent,Knaus2,Sch,adhesion,Kramer1,Gennes,Brener,Gert,Per1}.
In a recent series of papers Shrimali and Lopez-Pamies have presented an interesting new approach to
crack propagation in viscoelastic solids\cite{Lop1,Lop2,Lop3}. The theory is based on the assumption that
a slab of rubber fracture at a critical stretch that is independent of the stretch rate. 
Here we argue that the study is not as general as stated by the authors. 
In what follows we assume for simplicity linear viscoelasticity and neglect inertia effects.

\begin{figure}[!ht]
\includegraphics[width=0.4\textwidth]{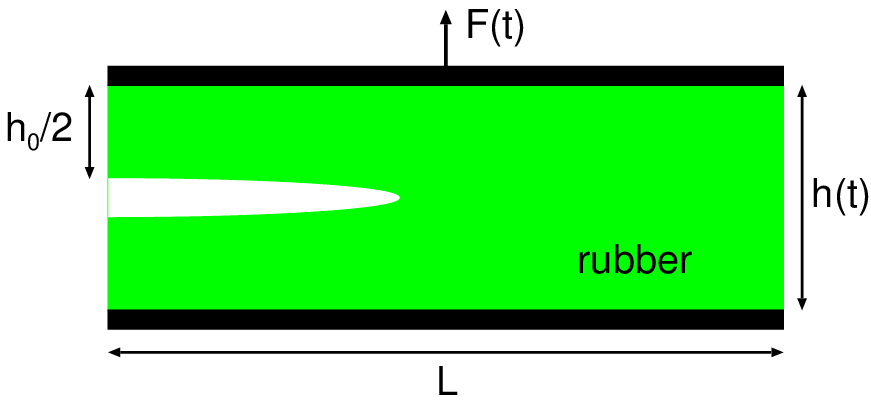}
\caption{\label{crack.eps}
Crack in a rubber sheet with clamped upper and lower edges. The height $h(t) = h_0+u(t)$ of the slab increases
with time (undeformed height is $h_0$). For an elastic solid (no viscoelasticity) the crack start to move
when the elongation strain $\epsilon = u(t)/h_0$ reach the critical value $(2 \gamma /E h_0)^{1/2}$ where $\gamma$
is the energy per unit surface area to break the bonds at the crack tip.}
\end{figure}

\vskip 0.2cm
{\bf A Dependency of the strain at fracture on the strain rate}
 
The theory of Shrimali and Lopez-Pamies is based on the assumption that fracture in ``pure shear'' 
(see Fig. \ref{crack.eps}) occur
at a critical stretch that is independent of the stretch rate. As support for this they cite several experimental studies where
this is indeed observed\cite{ex1,ex2}. However, the experimental studies are for a limited range of applied stretch rates,
and it is easy to show that it cannot be true in general. To show this consider
a slab of elastomer (of undeformed height $h_0$) elongated at a rate $\dot h$. We assume that the length $L$ of the slab is much bigger than the
height $h_0$ and that the crack tip is not close (on the scale of $h_0$) to any end of the slab.
For simplicity we assume linear viscoelasticity with the frequency-dependent modulus $E(\omega)$.
The modulus in the low-frequency (rubbery) region $E(0)=E_0$ is much smaller than the modulus $E_1 = E(\infty)$
in the high frequency (glassy) region. (Note that the rubbery region and the glassy region may be separated by 10 (or more)
decades in frequency.) If the elongation strain rate 
$\dot \epsilon = \dot h/h_0$ (or rather strain rate frequency $\omega = \dot \epsilon/\epsilon$)
is very low the rubber will effectively be in the rubbery state and the onset of crack
propagation will occur when the strain $\epsilon_0 = (2\gamma /h_0 E_0 )^{1/2}$ where $\gamma$ (denoted by $G_{\rm c}$ 
in Ref. \cite{Lop1}) is the energy per unit surface area
to breaks the bonds at the crack tip. However, if the strain rate is very high the fracture will occur almost
instantaneously and rubber will effectively be in the glassy state everywhere 
both with respect to the deformations resulting from the elongation 
and from the movement of the crack tip. Hence we can again apply the standard Griffith fracture criteria but now using the high
frequency modulus so that the critical (fracture) strain $\epsilon_1 = (2\gamma /h_0 E_1 )^{1/2}$ which is smaller than
the low strain rate fracture strain by a factor of $(E_0/E_1)^{1/2}$ which is typically in the range $0.1-0.01$ in practical cases.
This conclusion is consistent with the temperature-frequency relation valid for (simple) rubber materials\cite{WLF}
where high frequencies are equivalent to low temperatures. At temperatures below
the glass transition temperature the time scale of deformation is typically much higher than that of the
(Brownian motion) of the molecular chains of the rubber, and rubber compounds are hard and brittle like glass.

Experiments we have performed show that a critical stretch that is independent of the stretch rate is a special case 
which is mostly encountered (in a finite stretch rate range) in very tough viscoelastic materials\cite{Creton1,Cristiano}. 
We have performed few tests with the pure shear geometry but many tests in singe-edge notch geometry 
(where the critical stretch depends on notch length) and the stretch at break is markedly strain rate dependent, 
which may in part result from a strain rate dependency of the bond breaking in the process zone. 
See Ref. \cite{Creton1} and an earlier study on the fracture of very elastic polyurethanes\cite{Cristiano}. 

\vskip 0.2cm
{\bf B Definition of the fracture energy}
 
The fracture energy $G$ (denoted as $T_{\rm c}$ in Ref. \cite{Lop1}) is usually measured under conditions where the 
strain far from the crack tip is in the relaxed state. For example in some experiments the slab in Fig. \ref{crack.eps}
is first stretched without the crack and then kept in the stretched state for equilibrium to occur after which a 
crack is inserted using a razor blade. For this situation $G(v)$ is the energy to break the bonds at the crack tip plus
the viscoelastic energy dissipation close to the crack tip due to its motion, and will only depend on the crack tip speed $v$ (and the
temperature, see below). In this case $G(v)$ can be considered as a (useful) material parameter which can be tabulated
for different rubber compounds as done by Gent\cite{Gent} for three types of rubber [cis-polybutadiene (PB), styrene-butadiene 
copolymer (SB) and ethylene-propylenecopolymer(EPR)]. 

For a general situation where the strain field far from the crack is time dependent, as when the slab in Fig. \ref{crack.eps}
is elongated $h=h(t)$ in some arbitrary complex way, Shrimali and Lopez-Pamies {\it define} the fracture energy as
$G=-dW/dA$ where $dW/dA$ is the change in the total (stored and dissipated) 
deformation energy in the bulk with respect to added surface area $A$ of the pre-existing crack.
But using this definition $G$ will depend on the (time-dependent) loading conditions and is no longer
a material parameter but depend on the loading history. Shrimali and Lopez-Pamies only studied the onset of crack propagation
where $v=0$ but in the most general situation where the strain field far from the crack is time dependent, 
and the crack propagate with the speed $v$, the crack propagation
need to be studied theoretically (or experimentally) for each such case separately using, e.g., numerical methods (see Sec. F). 
The material input parameters for such studies are
the viscoelastic modulus and the energy to break the bonds per unit surface area $G_{\rm c} = \gamma$. 
(In general $\gamma$ depend on complex processes occurring in the crack tip process zone, which depend on the 
crack tip speed and the temperature, i.e., $\gamma = \gamma(v,T)$ is a function of the crack tip speed and the temperature.)

The theory proposed in Ref. \cite{Lop1} is stated to be valid only at the onset of
crack propagation. This case is not of very big interest in practical
applications since the (useful) toughness of rubber materials result from the fact that the crack propagation energy increases very strongly
with increasing crack tip speed, e.g., by a factor of $E_1/E_0$ when the stress field far from the crack is in the relaxed state.

Consider the system in Fig. \ref{crack.eps} exposed to time-dependent external forces. 
In Ref. \cite{Lop1} it was proposed that
the crack will start to move at the time $t$ when the condition
$$-{d W^{\rm Eq} \over d A }= G_{\rm c}\eqno(1)$$
is obeyed. Here $W^{\rm Eq}$ is the elastic energy stored in the system at equilibrium i.e. the elastic energy
after keeping the system which prevail at time $t$
for an infinite long time with fixed boundary conditions (as given by the boundary conditions at time $t$) 
on the part of the boundary that are not traction free. As support for this equation the author 
use the experimental observation that the fracture for the system shown in Fig. \ref{crack.eps} 
occur at a fixed strain independent of the strain rate.
However, we have argued above that this may be true only for low enough strain rates, and will 
break down for strain rates where the deformation frequency $\omega = \dot \epsilon/\epsilon = \dot u/u$ 
(see Fig. \ref{crack.eps}) occur in the glassy response region. 
This result in a violation of (1) as can be easily seen by the following argument:
Consider the system in Fig. \ref{crack.eps} and assume first the case of extremely slow stretch rate. 
In this case the material behaves as an elastic body with low-frequency elastic modulus $E_0$ 
and the energy balance would then require
$$ dW^{\rm Eq} = - G_{\rm c} dA\eqno(2)$$
where $dW^{\rm Eq}=dU$ is the change of elastic energy $dU=E_0 \epsilon_0^2 h_0 dA/2$ of the relaxed material.
Next consider extremely high stretch rate, i.e. a step change of the remote displacement. 
In this case the material is in the glassy region and will behave elastically with modulus 
$E_1$ and the energy balance will require that the total change of the elastic energy must balance the fracture energy
$$dW^{\rm Eq} +dW^{\rm NEq} =- G_{\rm c} dA\eqno(3)$$ 
Here the change in the  total elastic energy equal $dU=dW^{\rm Eq} +dW^{\rm NEq} =E_1 \epsilon_1^2 h_0 dA /2$.
The conditions (2) and (3) give the Griffith fracture results for $\epsilon_0$ ans $\epsilon_1$ quoted in Sec. A.
It is evident that Eq. (6) in Ref. \cite{Lop1} cannot hold true in all cases as the authors propose.

\vskip 0.2cm
{\bf C Velocity dependency of the crack propagation energy $G(v)$}
 
As we understand it the theory developed in Ref. \cite{Lop1,Lop2,Lop3} only address the onset of crack propagation. 
Therefore, the dependency of the fracture energy $G(v) = \gamma [1 + f(v)]$ 
on the crack-tip speed cannot be deduced from the theory.
When the crack is propagating at constant speed one has to consider 
the influence of non-conservative work of internal stresses to write down the 
correct energy balance equation. 
This was done in Ref. \cite{Carb} where Eqs. (6), (8), (9) and (10) give the change 
of the elastic energy, and of the non-conservative work of internal stresses, upon a small 
displacement of crack front both for the opening and closing crack, 
thus extending the Griffith criterion to viscoelastic cracks moving at constant speed.

Here we note that the classical experiments of Gent measured the $G(v)$ 
relation by peeling rubber sheets apart\cite{Gent}.
By performing measurements at different peeling speeds and temperatures and using
the temperature-velocity shifting procedure he could map out the full $G(v)$ (master) curve.
Measurements of this type cannot be analyzed using the theory of Shrimali and Lopez-Pamies. 

\begin{figure}[!ht]
\includegraphics[width=0.48\textwidth]{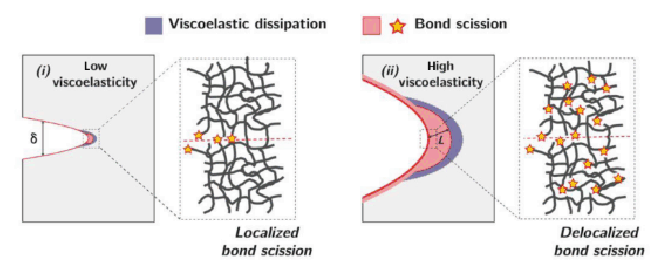}
\caption{\label{Tip.eps}
Schematic coupling between viscoelasticity (blue domain) and strand breakage (red domain) at the crack tip. The
enlarged region shows the occurrence of bond scission (yellow stars) in the elastomer network. Bond scission and viscoelastic
dissipation are strongly coupled, with a joint increase in bond scission 
and viscoelastic dissipation between the low viscoelasticity (i) and
large viscoelasticity regimes (ii). Adopted from Ref. \cite{Creton1}.}
\end{figure}

\vskip 0.2cm
{\bf D Crack tip process zone}
 
Shrimali and Lopez-Pamies claim that the advantage of their approach compared to earlier theories is that they do not need to
invoke a crack tip process zone. However, a crack tip process zone will occur in all real materials and it is important to study
its influence on the crack propagation. If one assume a priory that no crack tip process zone occur one cannot claim that
it will have no influence on the viscoelastic part of the crack propagation energy. Here we note that for cohesive crack
propagation the crack tip process zone can be very complex involving cavitation, formation of filaments (stringing) and a recent study
has shown that breaking of chemical bonds may occur far away from the crack tip in a region overlapping with the region where viscoelastic
energy dissipation occur, see Fig. \ref{Tip.eps} and Ref. \cite{Creton1}. 
Thus bond scission, far from being restricted to
a constant level near the crack plane, can be delocalized over hundreds of micrometers and
increase by a factor of 100, depending on the temperature and stretch rate, and the energy dissipated by covalent bond scission
accounts for a much larger fraction of the total fracture energy than was previously believed.
The situation may be less severe for adhesive crack propagation (crack propagation at the interface
between a flat rigid solid and an elastomer film adhering to the substrate) but even in these cases for weakly crosslinked 
elastomers (as in pressure sensitive adhesives) the crack tip process zone is very complex and spatially large\cite{Creton2}.

The existence of a (complex) process zone in rubber-like materials can be (indirectly) 
seen through the roughness profiles of the surfaces formed by the 
crack propagation\cite{Horst,Gorb}. This is valid for all real materials, not only rubber. The morphology of fracture surfaces 
reflect the complex processes occurring on different length scales close to the crack tip, and depend on the 
microstructure of the material. 

\vskip 0.2cm
{\bf E Delayed fracture}

Shrimali and Lopez-Pamies studied the so-called delayed fracture test\cite{Lop2}. 
In a typical delayed fracture test, a sheet of the
elastomer of interest containing a pre-existing crack is
subjected to a load that is applied rapidly over a very
short time interval $[0, t_0]$ and then held constant. 
Nucleation of fracture from the pre-existing crack occurs at
a critical time $t_{\rm c} > t_0$ , hence the name of the test.
Delayed fracture sometimes occurs also at fixed grips conditions 
(fixed stretch) for which $W^{\rm Eq}$ is time independent. 
This delayed fracture at constant stretch cannot be 
explained by the Shrimali and Lopez-Pamies theory.

\vskip 0.2cm
{\bf F Numerical approaches to viscoelastic crack propagation}
 
Shrimali and Lopez-Pamies performed a numerical study using the finite element method applied to
an elastomer with non-linear rheology. To determine when the crack start to move they used the criteria (1) [see
Eq. (6) in Ref. \cite{Lop1}] involving only $G_{\rm c}=\gamma$ and the equilibrium part of the stored elastic energy. 
In our opinion this approach may be valid only at the onset of crack propagation and for low enough stretch rate. 
An alternative more general approach, which can be used to obtain the full $G(v)$ curve, is to use
a discretized solids where the ``atoms'' are connected by (realistic) non-linear springs or bonds,
e.g using a non-linear Rouse-like model for the network chains\cite{Febbo}. Such atomistic
models have been used to study the $G(v)$ relation for crack propagation in silicone\cite{Si} 
and recently also for adhesive cracks involving viscoelastic solids in contact with a 
rigid flat substrate\cite{Mus}. An advantage with this approach is that it 
can also include inertia effects which are crucial
in some cases (see Ref. \cite{Si}).

%\vskip 0.2cm
%{\bf Acknowledgments:}
%We thank G. Heinrich and T. Tada for comments on the text.

\end{document}